\documentclass[epj]{svjour}
\usepackage{graphics}
\def\ra{\mbox{$\rightarrow$}} 
\def\rs{\mbox{$\sqrt{s}$}} 
\def\epem{\mbox{$\mathrm{e}^+\mathrm{e}^-$}}

\newcommand{\GEVc}{\mbox{$\mathrm{Ge\!V}/c$}}

\newcommand{\GEVcc}{\mbox{$\mathrm{Ge\!V}/{c^2}$}}
\newcommand{\ffbar}{\mbox{$\mathrm{f\overline{f}}$}}
\newcommand{\qqbar}{\mbox{$\mathrm{q\overline{q}}$}}
\newcommand{\qq}{\mbox{$\mathrm{qq}$}}
\newcommand{\gaga}{\mbox{$\gamma\gamma$}}
\newcommand{\ev}{\mbox{$\mathrm{e}\nu$}}

\newcommand{\lv}{\mbox{$\ell\nu$}}

\begin{document}
\title{Gaugino searches and constraints on supersymmetry}
\author{Paolo Azzurri
}                     
%
%
\institute{Scuola Normale Superiore, 
Piazza dei Cavalieri 7, 56100 Pisa, Italy.
}
\date{Received: date / Revised version: date}
%
\abstract{
The negative outcome of gaugino searches at LEP has been one of the 
most disappointing results for fans of supersymmetric models. 
In the framework of minimal supersymmetric models with GUT 
unification assumptions, the combined absence of supersymmetry 
and Higgs signals in the LEP data, 
sets stringent constraints on the models parameter
space, and lower limits on the mass of the lightest neutralino and 
of other supersymmetric particles. 
All limits are given at 95\% Confidence Level.
} 
\maketitle
\section{Gaugino searches}
\label{intro}
The production of gauginos in \epem collision has certainly been the most 
expected signal of supersymmetry (SUSY) during the last ten years. 
In Minimal Supersymmetric extensions of the Standard Model 
(MSSM models~\cite{mssm}), gauginos are the fermionic superpartners 
of Standard Model (SM) gauge and Higgs bosons, making up two charged 
gauginos (charginos $\chi_{1,2}^\pm$) 
and four neutral gauginos (neutralinos $\chi_{1,2,3,4}^0$). 
In models where SUSY breaking is mediated by gravity, and with 
unification assumptions at the GUT scale, the lightest neutralino 
($\chi_1^0$), is the natural lightest SUSY particle (LSP). 
When R-parity further conserves lepton and baryon number, the LSP 
is stable and weakly interacting, and is of interest as a possible 
cosmological component of non-baryonic cold dark matter.
The SUSY models considered in the following will assume a neutralino-LSP,
GUT unification in all sectors  
and conservation of R-parity, implying that SUSY particles can only be
produced in pairs.

\subsection{Chargino-pair searches}
\label{intro:1}
\begin{figure*}
\centerline{
\rotatebox{-90}{\resizebox{0.55\textwidth}{!}{%
  \includegraphics{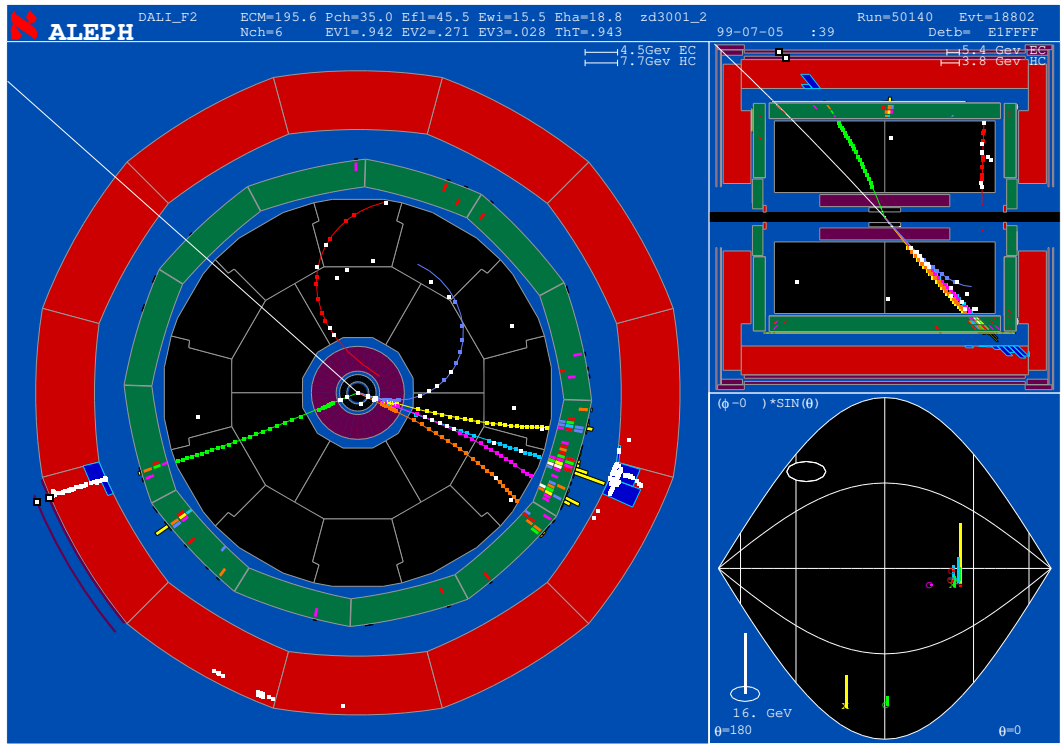}
}}}
\caption{A perfect chargino-pair candidate event recorded by ALEPH 
at \rs=196~GeV. The event display shows an isolated muon with 8.5~\GEVc\
momentum and an hadronic monojet with a 16~\GEVcc\ invariant mass. The 
total missing invariant mass in the event amounts to 148~\GEVcc, with a 
transverse missing energy component of 21~GeV. The possible Standard 
Model source of such an event is a W-pair decay 
WW$^\ast\ra\tau(\ra\mu\nu\nu)\nu\qqbar$, where the hadronic decay of
the second W is extremely off-shell.}
\label{fig:1}       
\end{figure*}

In chargino-pair productions 
each produced chargino will decay to a fermion-pair and a LSP according to
the chain
\begin{equation}
\label{eq:chap}
\epem\ra\chi^+\chi^-\ra\ffbar'\chi_1^0\ffbar'\chi_1^0 .
\end{equation}
The $\chi_1^0$ pair will carry away undetected 
energy while the two fermion-pairs
($\ffbar'\ffbar'$) can give rise to three main topologies 
with {\sl (i)} two leptons 
($\lv\lv$), {\sl (ii)} one lepton and two jets ($\qq\lv$), or 
{\sl (iii)} multi-jets ($\qq\qq$). 
The relative amount of visible and missing energy will depend 
mainly on the mass difference $\Delta M$ between the chargino and 
the LSP.
For large $\Delta M$ values the main SM backgrounds for the searches 
come from four-fermion events as W-pairs,Z-pairs, W$\ev$ and Zee. 
In the case of low $\Delta M$, main backgrounds arise from two photon
collisions as $\gaga\ra\tau\tau$, hadrons~\cite{gaug}.
In scenarios where $\Delta M$ is smaller that 3$\GEVcc$ the $\gaga$
background can only be reduced by requiring the presence of an initial
state radiation (ISR) photon with large transverse momentum, that will also 
ensure
the signal triggering, but at the expense of a signal selection efficiency 
lower than 3\%. Finally when $\Delta M$ is below the pion production
threshold, charginos can be detected as heavy stable charged 
particles~\cite{chad}.
A very nice chargino-pair candidate is shown in fig.~\ref{fig:1}.

\subsection{Neutralino-pair searches}
\label{intro:2}
Signal topologies arising from pair productions of heavy neutralinos can be 
quite different due to a variety of decay channels and to possible 
cascade processes. The main searched topology arises from 
\begin{equation}
\label{eq:neu1}
\epem\ra\chi^0_i\chi^0_1\ra\ffbar\chi_1^0\chi_1^0
\end{equation}
and gives final states with acoplanar jets and acoplanar leptons. 
When two fermion pairs are produced as from
\begin{equation}
\label{eq:neu2}
\epem\ra\chi^0_i\chi^0_j\ra\ffbar\ffbar\chi_1^0\chi_1^0
\end{equation}
the searched final states are multi-leptons or multi-jets with missing
energy. Finally to cope also with possible radiative 
$\chi_i^0\ra\chi_j^0\gamma$ decays signals as 
\begin{equation}
\label{eq:neu3}
\epem\ra\chi^0_i\chi^0_j\ra\ffbar\gamma\chi_1^0\chi_1^0
\end{equation}
are also searched for, with jets, photons and missing energy.
Again different searches are performed according to possible
$\Delta M$ mass differences between the heavy neutralinos 
and the $\chi_1^0$-LSP, with SM backgrounds arising mainly from 
four-fermion processes at large $\Delta M$, and from 
two-photon collisions at small $\Delta M$~\cite{gaug}.

%

\section{Lower LSP mass limits}
\label{sec:1}
In the LEP data up to \rs=209~GeV none of the SUSY searches has revealed 
an excess of signal events over the expected SM processes. Therefore upper 
limits on gaugino pair production cross sections have been set 
at the level of less than a pb 
for all main topologies, when kinematically accessible.
These cross section limits can be translated in exclusion domains in 
the constrained MSSM parameter space ($\tan\beta,\mu,M_2,m_0$), 
and from here in lower mass limits 
on the LSP and other SUSY particles.

\subsection{Limits with heavy sfermions}
\label{lsp:1}
When sfermions are heavier than electroweak bosons they do not interfere
with the gaugino production and decay. In this scenario the expected 
gaugino production cross sections are in the 1-10~pb range. The 
excluded parameter space corresponds roughly to domains where the chargino
mass is within the kinematic production limit,
{\it i.e.} $m(\chi^+)\leq \rs/2\approx $100~\GEVcc.
The corresponding limit on the LSP mass can be set at 
$m(\chi^0)>39~\GEVcc$, 
at $\tan\beta\approx 1$~\cite{gaug}.

\subsection{Limits with light sfermions}
\label{lsp:2}
The effects of light sfermions (with a GUT unified mass $m_0$) are 
significant both in the production and decay of gauginos. 
Using dedicated searches of different decay topologies and direct 
sfermion searches, the MSSM parameter space can, however, be covered 
in a way similar to the case of heavy sfermions, and the LSP limit 
still holds at $m(\chi^+)>39~\GEVcc$, 
with $\tan\beta\approx 1$~\cite{gaug}.

\subsection{Exclusions from Higgs boson searches}
\label{lsp:3}
The negative outcome of Higgs boson searches~\cite{higgs}
allows to exclude MSSM regions with $\tan\beta<2-3$, 
bringing the LSP mass limit 
at $m(\chi^0)>50~\GEVcc$ with $\tan\beta\approx 3$ in the case of heavy
sfermions, and at $m(\chi^0)>45~\GEVcc$ with $\tan\beta\geq 10$ 
in the case of light sfermions, where possible sneutrino-chargino 
mass degeneracies limit the exclusion reach~\cite{gaug}.

\subsection{Effects of sfermion mixing}
\label{lsp:4}
The possible effects of mixing in the third family of sfermions
($\tilde{\tau},\tilde{\rm b},\tilde{\rm t}$) have also been considered,
and open more mass degenerate scenarios where standard searches lose 
their sensitivity. The sensitivity is recovered with more 
dedicated searches, in particular for light $\tilde{\tau}_1$ signals in
gaugino cascade decays, and the final LSP mass limits are only 
slightly deteriorated to $m(\chi^0)>37~\GEVcc$ at $\tan\beta\approx 1$,
and are maintained at $m(\chi^0)>45~\GEVcc$ at large $\tan\beta$,
including Higgs exclusions~\cite{gaug}.

A summary of LSP mass limits in the constrained MSSM is shown 
in table~\ref{tab:1}, with different hypothesis and constraints.

\begin{table}[h]
\caption{Summary of lower mass limits on the LSP}
\label{tab:1}       
\begin{tabular}{c|cc}
\hline\noalign{\smallskip}
$m(\chi_1^0)$ limit & any $\tan\beta$ & \mbox{with Higgs constraints}  \\
\noalign{\smallskip}\hline\noalign{\smallskip}
heavy sfermions & $>$39~\GEVcc & $>$50~\GEVcc \\
any $m_0$       & $>$39~\GEVcc & $>$45~\GEVcc \\
with $\tilde{\rm f}$ mixing     & $>$37~\GEVcc & $>$45~\GEVcc \\
\noalign{\smallskip}\hline
\end{tabular}
\end{table}

\subsection{Limits in minimal supergravity}
\label{lsp:5}
\begin{figure}[h]
\resizebox{0.55\textwidth}{!}{%
  \includegraphics{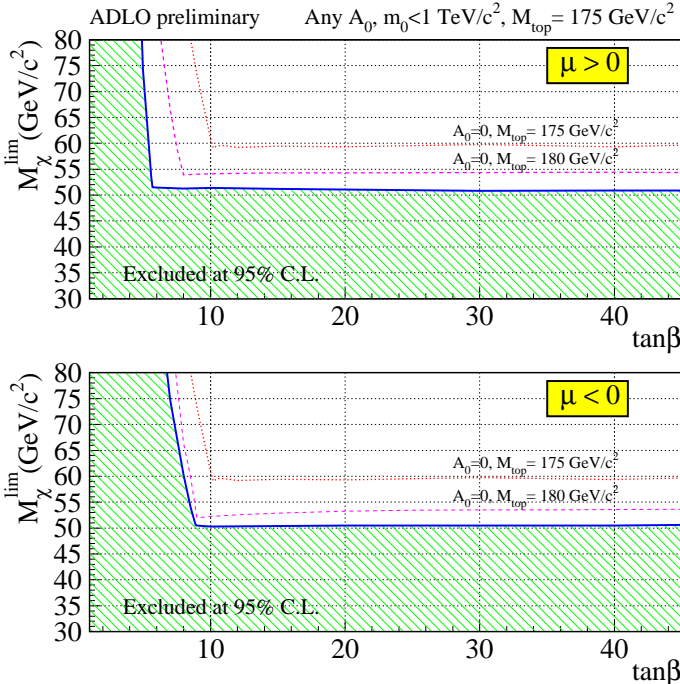}
}
\caption{Combined LSP lower mass limits in mSUGRA, as a function of
$\tan\beta$ for $\mu>0$ (top), and $\mu<0$ (bottom).}
\label{fig:2}       
\end{figure}

In the framework of minimal supergravity (mSUGRA) 
the MSSM parameters are further reduced 
and the LSP mass limits, with light sfermions, mixing and Higgs constraints,
improves to  $m(\chi^0)>50-60~\GEVcc$, depending on the top quark
mass~\cite{msugra}, as shown in fig.~\ref{fig:2}.

%
\subsection{Radiative corrections}
\label{lsp:6}
Radiative corrections to the gaugino unification relation 
$\frac{M_1}{M_2}=\frac53\tan^2\theta^2_W\approx\frac12$ 
and to the tree level gaugino masses
are at the level of 5\% and are not taken into account in the derivation
of the mass limits. The proper inclusion of these corrections could
deteriorate the LSP mass limits at the level of 1-2~\GEVcc, 
that sets the level of precision of the quoted results.
\section{Lower mass limits on other SUSY particles }
\label{sec:2}
Using the same exclusions of the constrained MSSM parameter space 
used to derive the LSP limits, lower mass limits on other 
SUSY particles can be derived, and are shown in table~\ref{tab:2},
again under the assumption of unified light sfermions and using 
the constraints from Higgs searches~\cite{gaug}.

\begin{table}[h]
\caption{Other lower mass limits on SUSY particles}
\label{tab:2}       
\begin{tabular}{c|c}
\hline\noalign{\smallskip}
SUSY particle & 
lower mass limit with \\
&  any $m_0$ and Higgs constraints \\
\noalign{\smallskip}\hline\noalign{\smallskip}
chargino & $m(\chi^+)>90-100~\GEVcc $ \\
sneutrino & $m(\tilde{\nu})>85-95~\GEVcc $\\
right selectron &  $m(\tilde{\rm e}_R)>75-95~\GEVcc $ \\ 
left selectron  & $m(\tilde{\rm e}_L)>110~\GEVcc $  \\
\noalign{\smallskip}\hline
\end{tabular}
\end{table}

%
%

\end{document}